\documentclass{psp-book9x6}
\usepackage{psp-book-har}           
\usepackage{subfigure}

\begin{document}

\setcounter{page}{1}

\chapter[Controlling the fluid-fluid mixing-demixing phase transition with electric fields]{Controlling the fluid-fluid mixing-demixing phase transition with electric fields
\label{ch_tsori}}

\chapauth{Jennifer Galanis, Sela Samin, and Yoav Tsori\chapaff{Department of Chemical Engineering, Ben Gurion University of the Negev, Beer-Sheva, 84105, Israel\\
tsori@bgu.ac.il}}

Various properties of a material (viscosity, refractive index, etc.) can dramatically change after a phase transition. Investigating exactly when and how this change occurs potentially elucidates the dominant molecular forces working within the material, thus fueling significant scientific interest. Similarly, the analysis of idealized molecular models provides a framework for predicting the behavior of real, and often more complicated, materials. Understanding phase behavior is not just limited to fundamental research, as the ability to switch on or off various desirable properties with a simple ``turn of the knob'' can lead to interesting technological applications. 

Traditionally, intrinsic thermodynamic variables, such as temperature or pressure, act as the ``knobs'' that control phases. However, the application of \emph{external} fields such as gravitational, magnetic, and electric fields, can play an equally vital roll since they often offer a reversible, easily tunable, and localized method of control. The necessary key for practical use resides in a strong coupling between the field and a material property, for example the applied electric field with the material's dielectric response function. If such a strong coupling can be obtained, large changes to the phase diagram can occur even with a weak field.

In this chapter, we will specifically focus on theoretical advancements for how electric fields induce a mixing-demixing phase transition between two dielectric fluids (oils), presenting results from both equilibrium and dynamics. Moreover, we will highlight how strong coupling between the field and dielectric function naturally occurs with \emph{nonuniform} electric fields originating from small curved charged objects.

\section{Equilibrium phase behavior}\label{tsori_statics}
Using a mean-field approach, we consider a binary mixture of two fluids, $A$ and $B$, in an electric field $\mathbf{E}$, and write the total free energy $\mathcal{F}$ for a volume $V$ as
\begin{equation}
\label{eq_FE_tot}
\mathcal{F} = \int_V \left( f_{\mathrm{m}} + f_{\mathrm{e}} +
f_{\mathrm{i}} \right)
\mathrm{d}V
\end{equation}
where $f_{\mathrm{m}}$, $f_{\mathrm{e}}$, and $f_{\mathrm{i}}$ are the free energy densities for mixing, electrostatics, and fluid-fluid interfaces, respectively. 

The fluids, in the absence of an electric field, can mix or demix due to a competition between entropy and enthalpy, where temperature $T$ adjusts the relative balance. For concreteness, we consider the Bragg-Williams form
\begin{equation}
f_{\mathrm{m}} = \frac{kT}{v_0} \Big[ \phi \ln \phi + (1-\phi)\ln (1-\phi)
+\chi \phi (1-\phi) \label{eq_FE_ln} \Big]
\end{equation}
where $k$ is Boltzmann's constant, $v_0$ is the molecular volume of both components, $\phi$ such that $0<\phi<1$ is the volume fraction of component $A$, and $\chi\sim 1/T$ is the Flory interaction parameter~\cite{safran_book}. The first two terms account for entropy, while the third term accounts for the energy of mixing.

Equation~\ref{eq_FE_ln} gives rise to an upper critical solution temperature-type phase diagram. The shape of $f_{\mathrm{m}}(\phi)$ transitions from a single minimum at $\phi=1/2$ to a double minimum curve as $T$ shifts from above to below the critical temperature $T_c$. By using the well-known double tangent construction, we obtain the two binodal compositions $\phi_b$ for each temperature $T<T_c$. These compositions mark the mixing-demixing boundary in the $\phi-T$ plane. Specifically, fluids demix when $\phi$ is located between the two values of $\phi_b$, or more simply stated ``under the binodal curve''. Finally, the phase diagram terminates at the mixture's critical point $(\phi_c,\chi_c)=(1/2,2)$. 

In the following, we will employ a simplified form of eq.~\ref{eq_FE_ln}, namely the Landau expansion of the mixing energy around $\phi=\phi_c$:
\begin{equation}
f_{\mathrm{m}} \approx \frac{kT}{v_0}\left[
(2-\chi)\left(\phi-\frac{1}{2}\right)^2+ 
\frac{4}{3}\left(\phi-\frac{1}{2}\right)^4 + const. \right] \label{eq_FE_Landau}
\end{equation}
Note that the quadratic term in the expansion changes sign at the critical value $\chi_c=2$.

For electrostatics, the free energy density $f_{\mathrm{e}}$ is given by
\begin{equation}
\label{eq_FE_es}
f_{\mathrm{e}} =
\pm\frac{1}{2}\varepsilon_0\varepsilon(\phi)|\nabla\psi|^2
\end{equation}
where $\varepsilon_0$ is the vacuum permittivity, $\varepsilon(\phi)$ is the relative dielectric constant of the mixture, and $\psi$ is the electrostatic potential (${\bf E}=-\nabla\psi$). The positive (negative) sign corresponds to constant charge (potential) boundary conditions. The relation $\varepsilon(\phi)$ can, in fact, be a complicated function. Using the simplest approximation, we assume a linear relation, $\varepsilon(\phi) = (\varepsilon_A-\varepsilon_B)\phi + \varepsilon_B$, where $\varepsilon_A$ and $\varepsilon_B$ are the dielectric constants for pure fluids $A$ and $B$, respectively.

When considering the structure of the interface between phases, $f_{\mathrm{i}}$ is required. This term accounts for the energetic cost of composition gradients, and is given by~\cite{safran_book}:
\begin{equation}
f_{\mathrm{i}} = \frac{kT}{2v_0} \chi\lambda^2|\nabla\phi|^2
\end{equation}
where $\lambda$ is a constant characterizing the interface width. Notice that $f_{\rm i}$ vanishes when the composition is uniform.

To determine the equilibrium state in the presence of a field, we minimize $\mathcal{F}$ with respect to $\phi$ and $\psi$ using calculus of variations and obtain the following Euler-Lagrange equations
\begin{eqnarray}
\frac{\delta \mathcal{F}}{\delta\psi} &=&
\nabla\cdot\left[\varepsilon_0\varepsilon(\phi)\nabla\psi\right] = 0
\label{eq_EL_LandauGinzburg_ele} \\
\frac{\delta \mathcal{F}}{\delta\phi} &=& \frac{kT}{v_0}\left[
\left(4-2\chi\right)\left(\phi - \frac{1}{2}\right) +
\frac{16}{3}\left(\phi-\frac{1}{2}\right)^3-\chi\lambda^2 \nabla^2\phi
\nonumber \right] \\ && -
\frac{\varepsilon_0}{2}\frac{\mathrm{d}\varepsilon(\phi)}{\mathrm{d}\phi}
|\nabla\psi|^2  = \mu \label{eq_EL_LandauGinzburg}
\end{eqnarray}
The first equation is naturally Laplace's equation (Gauss's law) for the potential $\psi$, while the second equation gives the composition distribution $\phi$. Finally, $\varepsilon(\phi)$ couples these two equations---in our case, $\mathrm{d}\varepsilon(\phi)/\mathrm{d}\phi$ is constant.

The Lagrange multiplier $\mu$ in eq.~\ref{eq_EL_LandauGinzburg} differentiates between open and closed systems. For a closed system (canonical ensemble), $\mu$ is adjusted to satisfy the mass conservation constraint: $\langle \phi \rangle = \phi_0$, where $\phi_0$ is the average composition. A closed system, whose volume increases to infinity, can be related to an open system in contact with a material reservoir (grand canonical ensemble). At this infinite size limit, the mass conservation constraint can be approximated as $\mu=\mu_0(\phi_0)$, basically the chemical potential that corresponds to the reservoir composition $\phi_0$. The distinction between open and closed systems confers differences in phase behavior, as will be discussed below.

Before continuing to nonuniform fields, we will briefly review changes in the phase diagram with uniform fields. The effect of a uniform electric field, $\mathbf{E}_0$, on the mixture phase behavior was first studied by Landau and Lifshitz~\cite{landau_problem} and later by Onuki~\cite{onuki1995}. In the Landau theory, expansion of the electrostatic free energy leaves only a term proportional to $(\phi-\phi_c)^2$, which combines with the quadratic term in eq.~\ref{eq_FE_Landau} to renormalize the critical temperature and the entire binodal curve. The theory predicts a critical temperature shift: $\Delta T_c =v_0\varepsilon_0\varepsilon ''\mathbf{E}_0^2/(4k)$ that is controlled by two free parameters $\mathbf{E}_0$ and $\varepsilon ''=\mathrm{d}^2\varepsilon(\phi)/\mathrm{d}\phi^2$. If $\varepsilon ''$ is greater (less) than zero, $T_c$ increases (decreases) and the electric field effect is that of demixing (mixing).

Since $v_0$ is small in simple fluids, large electric fields are required to see an effect. For the typical maximal fields used in experiments ($\cong10^7$V/m), the predicted $\Delta T_c$ is extraordinarily small, on  the order of miliKelvins. Experiments in low molecular weight binary mixtures agree with the theory on the magnitude of $\Delta T_c$, but yield conflicting results on the sign or direction of the shift ~\cite{debye1965,orzechowski1999}. A more rigorous discussion about the effects of uniform electric fields is given in Ref.~\cite{tsori2009}.

Nonuniform electric fields, however, generate different results and can alter the mixing-demixing phase diagram considerably, compared to uniform fields of the same magnitude~\cite{tsori2004,marcus2008,samin2009}. Large field gradients occur naturally in systems like microfluidic and nanoscale devices due to their small size and complex geometry. Detailed investigations of the phase transition have been conducted with three simple yet fundamental shapes---wedge, sphere, and cylinder. Analogous results occur between these shapes; therefore due to space constraints, we focus on a closed system consisting of two concentric cylinders with radii $R_1$ and $R_2$, where $R_2 \rightarrow \infty$ produces an open system. We impose cylindrical symmetry such that $\phi=\phi(r)$ and $\psi=\psi(r)$, where $r$ is the distance from the inner cylinder's center. Furthermore, the prescribed charge density $\sigma$ per unit area on the inner cylinder allows integration of Gauss's law to obtain an explicit expression for the electric field: $\mathbf{E}(r) = \sigma R_1/ (\varepsilon_0\varepsilon(\phi)r)\mathbf{\hat{r}}$. Combining this result with $\mathbf{E}=-\nabla\psi$ in eq.~\ref{eq_EL_LandauGinzburg}, we obtain a single equation determining the composition profile $\phi(r)$:
\begin{equation}
\mu = \frac{kT}{v_0}\left[ \frac{\partial f_{\mathrm{m}}}{\partial\phi} -\chi\lambda^2 \nabla^2\phi - \chi M \frac{\mathrm{d}\varepsilon(\phi)/\mathrm{d}\phi}{\varepsilon(\phi)^2}\tilde{r}^{-2} \right]
\label{eq_EL_cyl}
\end{equation}
where $M = \sigma^2 v_0/(4kT_c\varepsilon_0)$ is the dimensionless field, and $\tilde{r}\equiv r/R_1$ is the scaled distance.

Inspection of eq.~\ref{eq_EL_cyl} shows two important differences from the case of uniform fields. First, the equation does not require a nonzero $\varepsilon''$ to shift the phase diagram. This holds irrespective of the three fundamental geometries. Recall that we, in fact, assumed a linear dependence for $\varepsilon(\phi)$. Such a simple constitutive relation is insufficient for changing the phase diagram in a uniform electric field, resulting in $\Delta T_c=0$.

\begin{figure}[tb]%
\begin{center}
\subfigure[\label{fig_f}]{\includegraphics[keepaspectratio=true,width=0.32\textwidth]{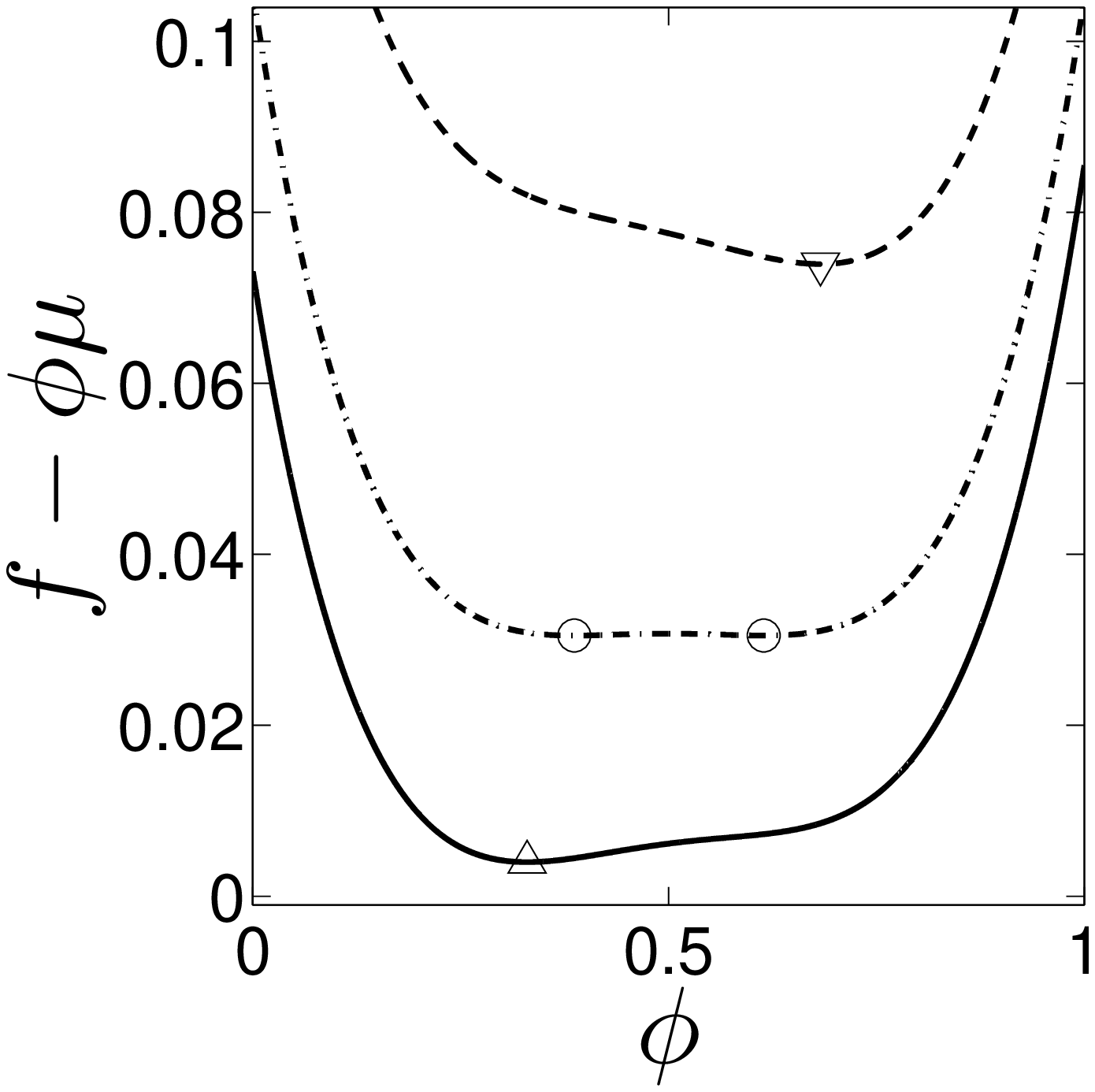}}
\subfigure[\label{fig_phi_r}]{\includegraphics[keepaspectratio=true,width=0.32\textwidth]{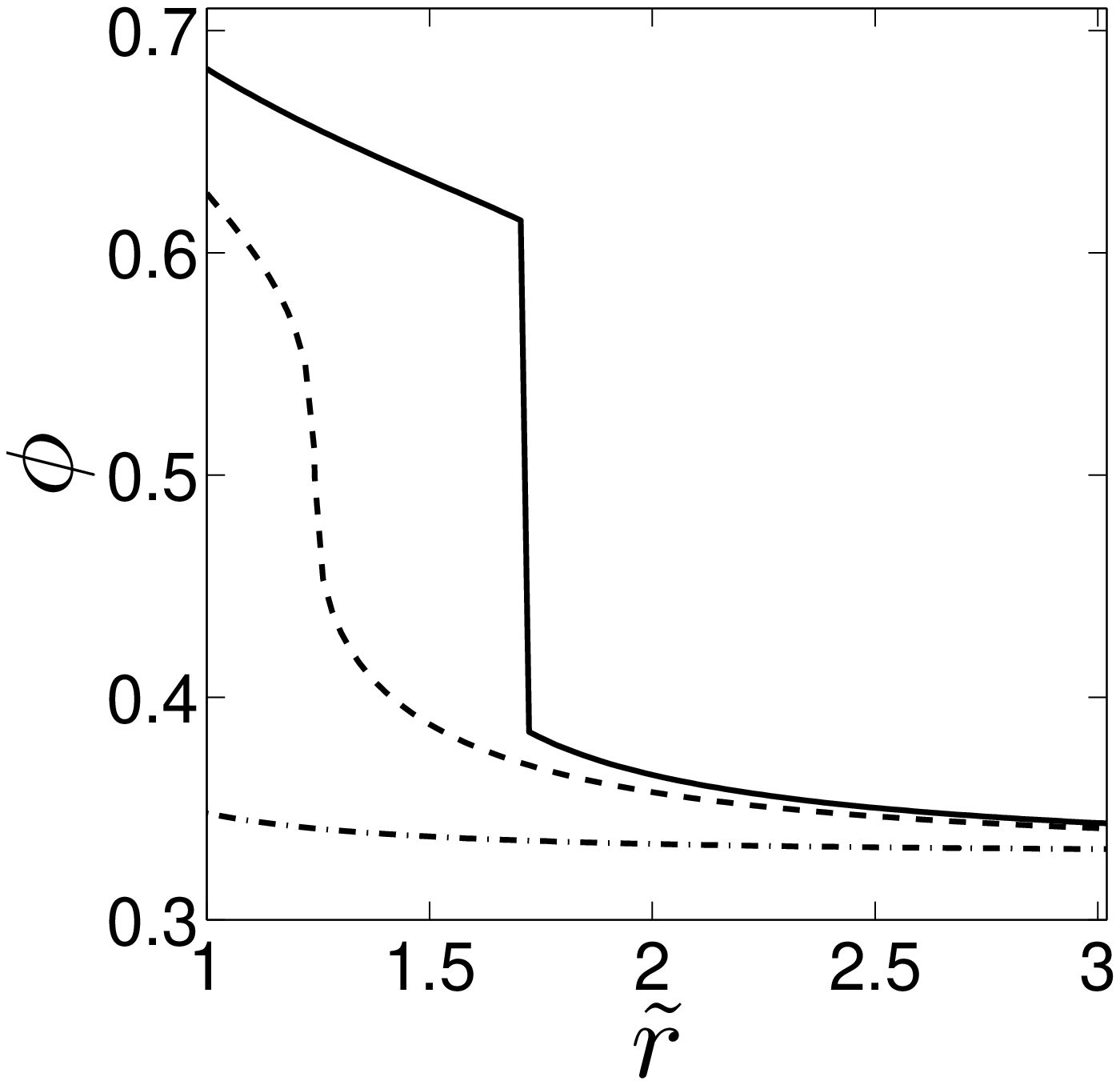}}%
\subfigure[\label{fig_varM}]{\includegraphics[keepaspectratio=true,width=0.32\textwidth]{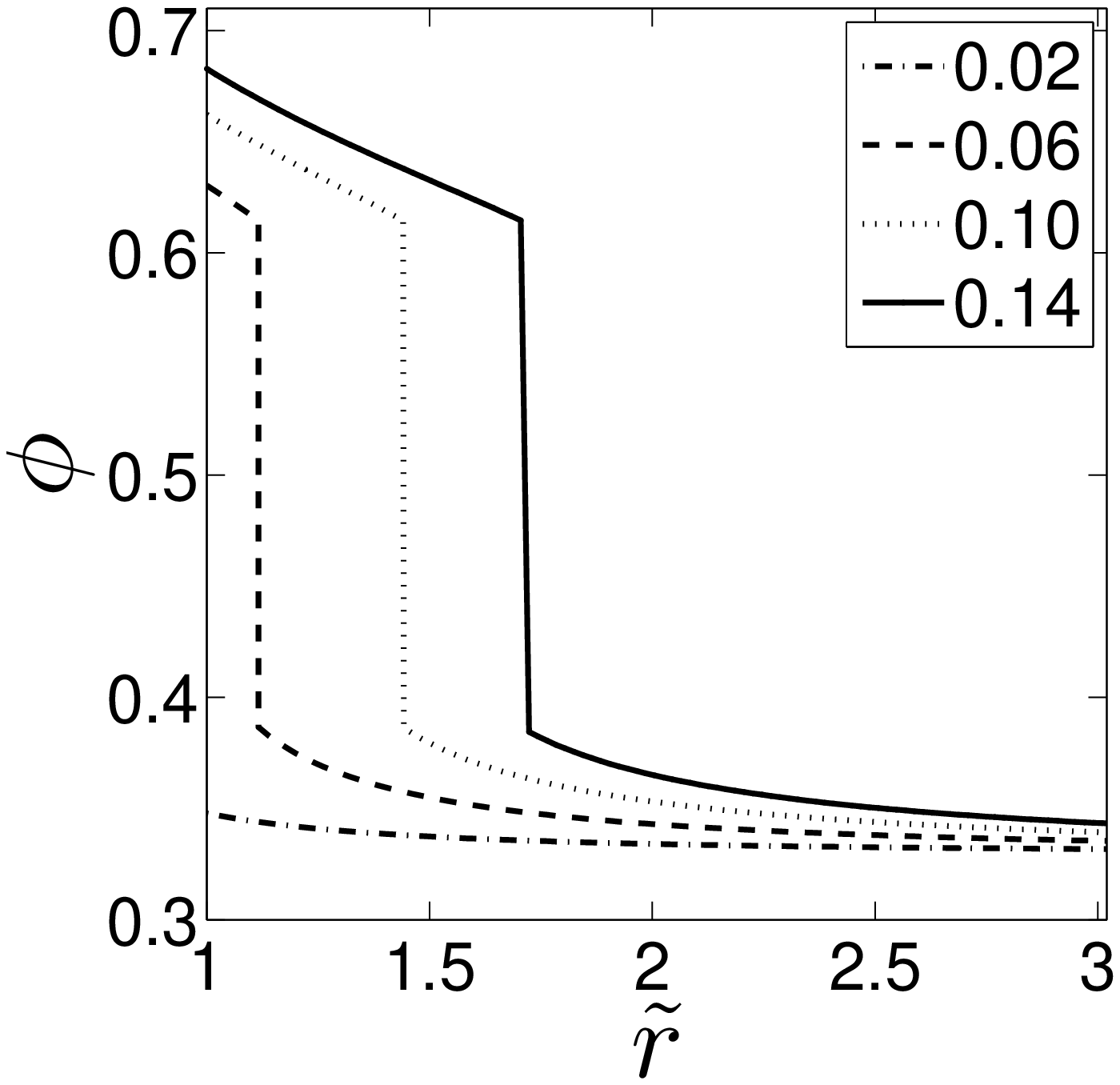}}%
  \caption{The free energy density $f(\phi,r)$ for a charged cylinder open system. (a) $f(\phi,r)-\phi\mu$ versus $\phi$ at distance $r=r_1$ (dashed line), $r_i$ (dash-doted line), and $\infty$ (solid line) for $\phi_0 = 0.33$, $T/T_c = 0.98$, and $M=0.14$. Symbols mark minima for each curve. (b) $\phi(r)$ versus dimensionless distance $\tilde{r}$. Solid line is data from (a). Dash-dotted line has same $\phi_0$ and $T$ as (a) but with $M=0.02$. Dashed line has same $\phi_0$ and $M$ as (a), but with $T/T_c=0.995$. (c) $\phi(r)$ versus $r$ for various values of $M$ with $\phi$ and $T$ as in (a).}%
\label{F_Fme}%
\end{center}
\end{figure}

Second, the nonuniform electric field imposes a nonuniform ``pull'' on the fluid mixture, manifesting as an $r$-dependent total free energy density $f(\phi,r)=f_{\mathrm{m}}+f_{\mathrm{e}}+f_{\mathrm{i}}$. We will consider, for illustrative clarity, a fluid-fluid interface that is infinitely thin by specifically defining a vanishing interfacial term $\lambda = 0$, which sets $f_{\mathrm{i}}=0$. With this simplification, the behavior of $f$ in an open system can be conceptualized as a competition between $f_{\mathrm{m}}$ and $f_{\mathrm{e}}$. As $r\to\infty$, the electric field is weak, $f_{\mathrm{e}}\to 0$, and $f\approx f_{\mathrm{m}}$ governs fluid behavior. The solid line in Fig.~\ref{fig_f} shows a typical example of $f(\phi,r)$ at a large value of $r$ using $\phi_0=0.33$, $T/T_c=0.98$, and $M=0.14$. The minimum of $f(\phi,r)$, marked by a symbol, gives the value of $\phi(r)$ as $r\to\infty$, which in this case is $0.33$. At the other distance extreme, $r=R_1$, the electric field is the strongest, and the dashed line in Fig.~\ref{fig_f} shows the resulting $f(\phi,r)$. Note the dramatic difference in the value of $\phi(r)$ when the value of $r$ is small ($R_1$) versus large. 

By finding the minima of $f$ for all $r$, it is possible to construct the full concentration profile $\phi(r)$, where the solid line in Fig.~\ref{fig_phi_r} corresponds to the data from Fig.~\ref{fig_f}. Whether or not a phase transition occurs resides in how the minimum of $f(\phi)$ changes as $r$ varies between the two distance extremes. Specifically, if there exists an $r = r_i$ where $f$ contains \emph{two} minima (see dash-dotted line in Fig.~\ref{fig_f}), then $r_i$ marks the interface between the two fluids. Figure~\ref{fig_phi_r} illustrates how the two minima in $f(r_i)$ translates into a discontinuity at $\phi(r_i)$, thereby creating a distinct boundary between the two phases.

Not all applied fields, however, induce phase separation. Figure~\ref{fig_phi_r} illustrates two such examples, where the dash-dotted line shows the same $\phi_0$ and $T$ with a smaller $M$ and the dashed line shows the same $\phi_0$ and $M$ with a higher $T$. In both cases, $f$ contains a single minimum for all $r$, resulting in a smooth $\phi(r)$ profile. Despite the absence of a phase transition in these examples, the field still produces an effect, as the more polar (higher $\varepsilon$) fluid accumulates near the high electric field. This phenomena can be thought of as the molecular version of the ``dielectric rise'' effect due to a dielectrophoretic force.

\begin{figure}[tb]%
\begin{center}
\subfigure[\label{fig_binodal}]{\includegraphics[keepaspectratio=true,width=0.32\textwidth]{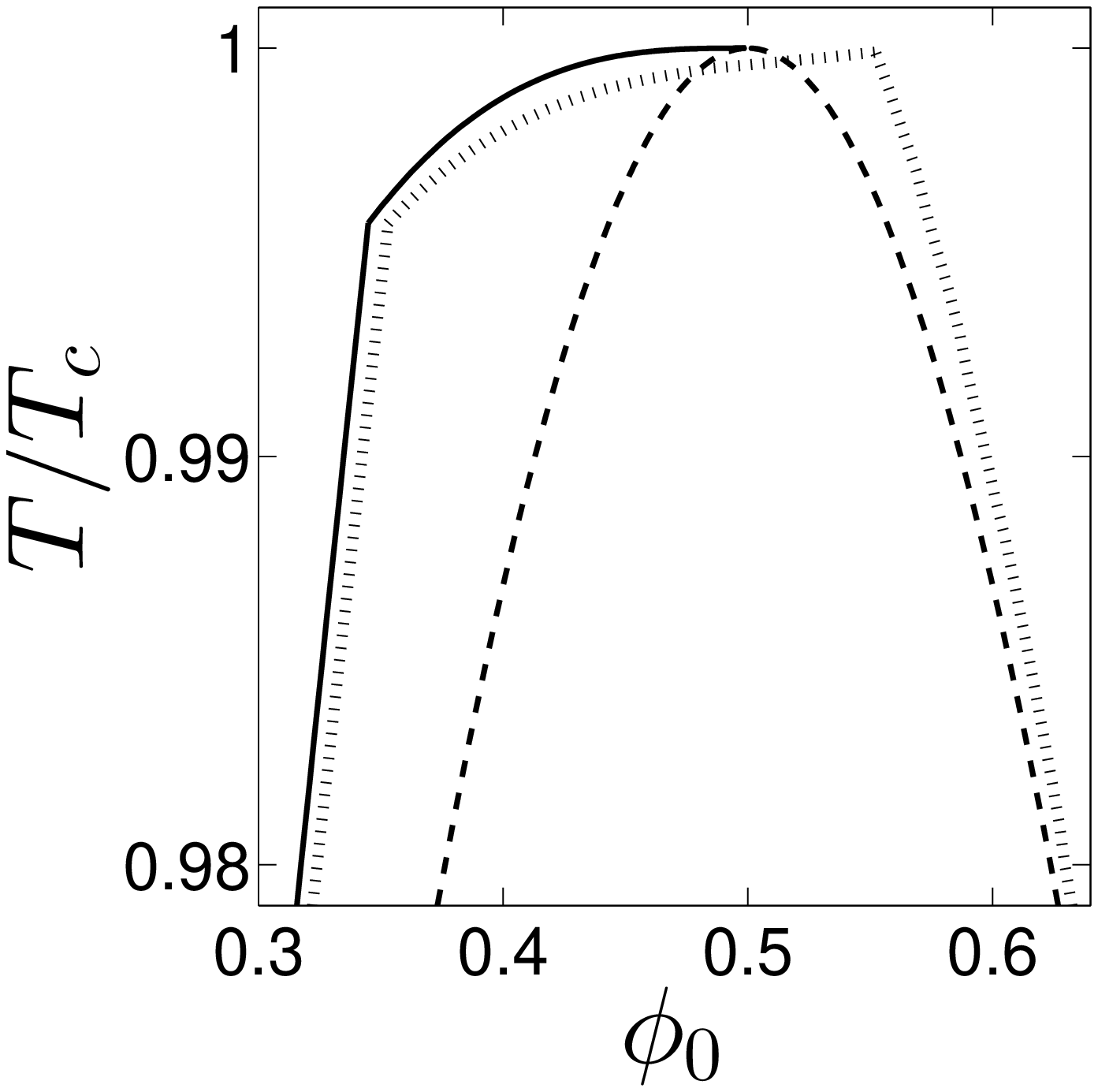}}
\subfigure[\label{fig_sigma}]{\includegraphics[keepaspectratio=true,width=0.32\textwidth]{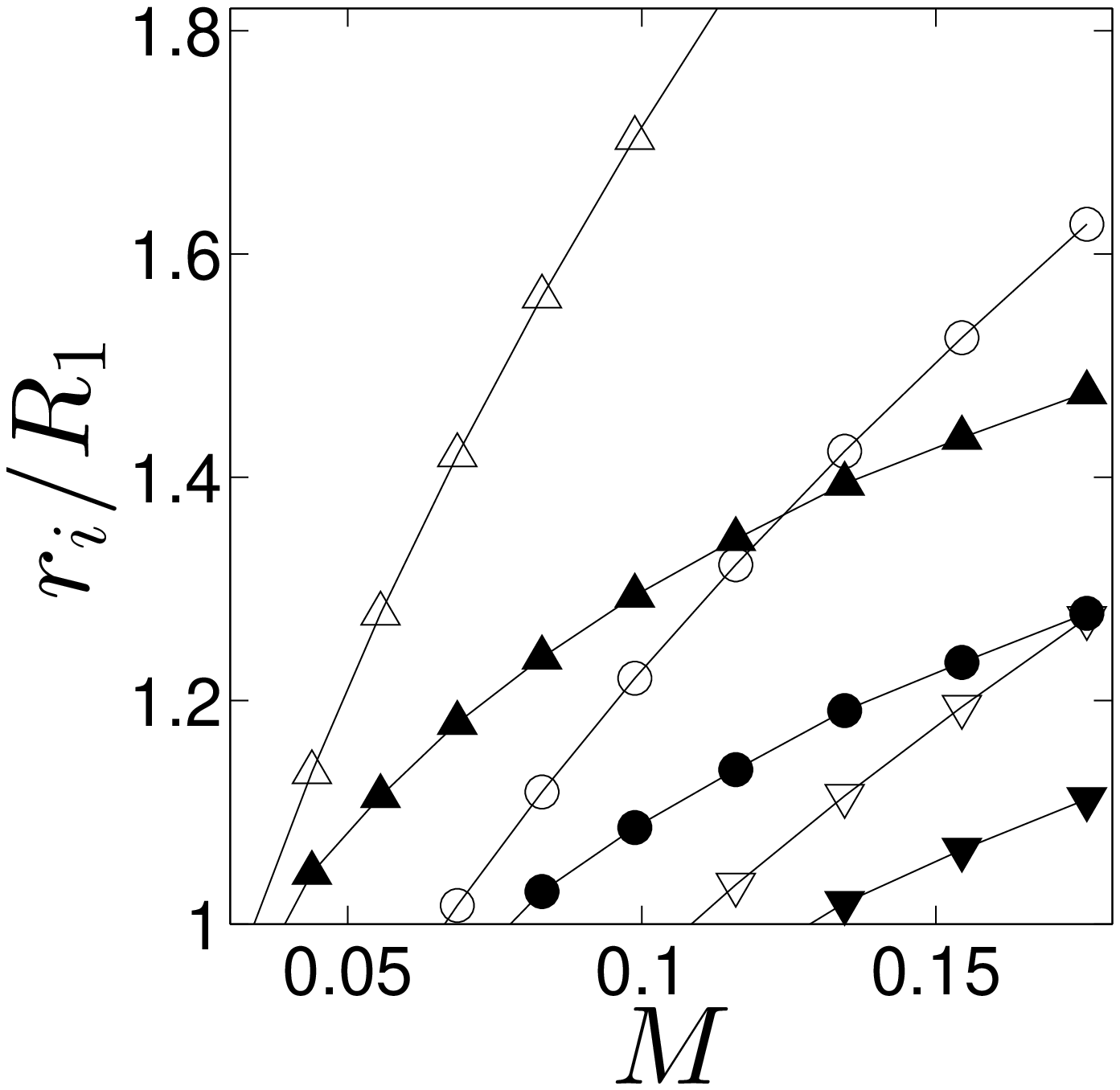}}
\subfigure[\label{fig_T}]{\includegraphics[keepaspectratio=true,width=0.32\textwidth]{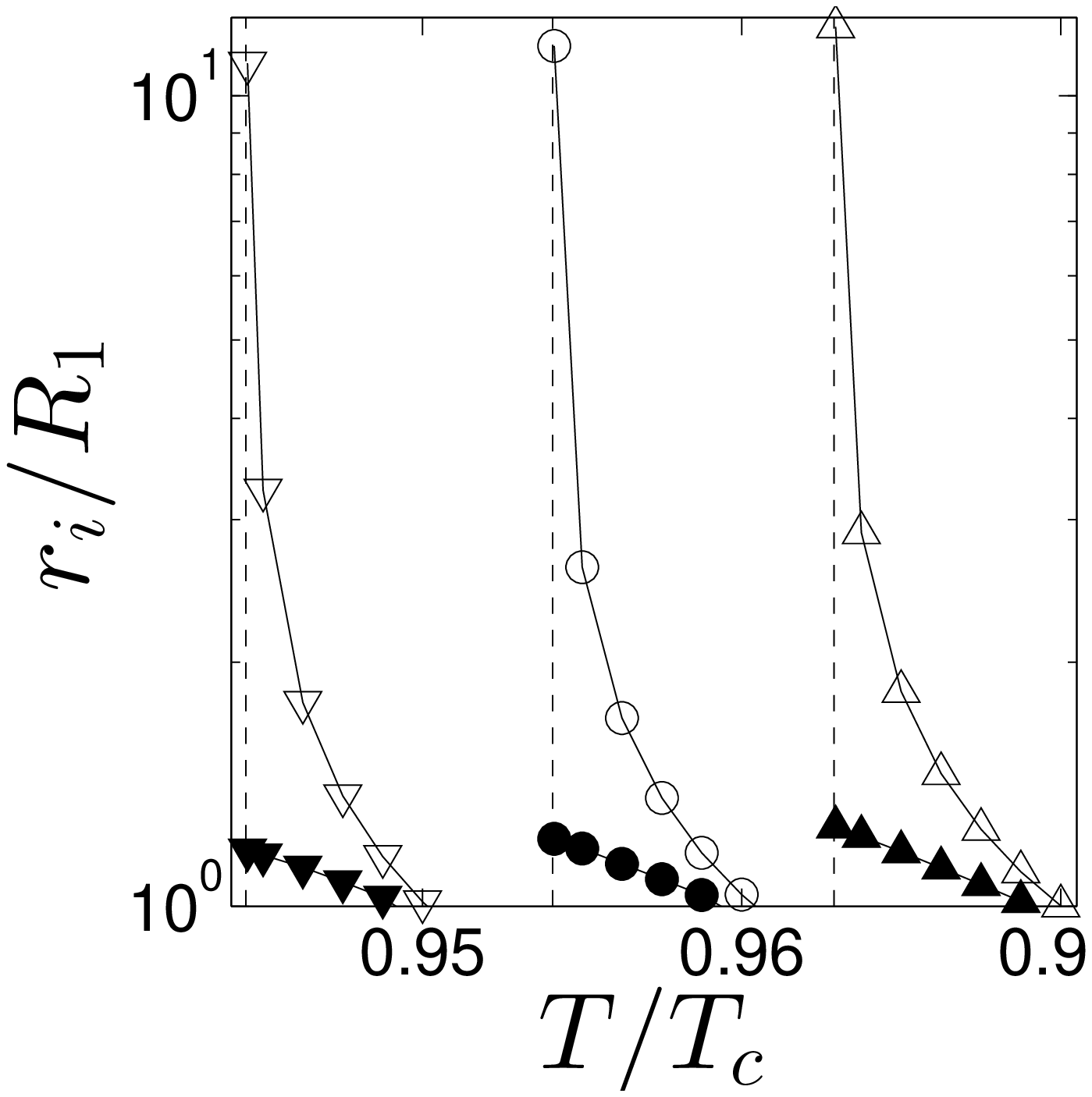}}%
  \caption{Behavior of the fluid-fluid interface $r_i$. (a) The stability curves in $\phi_0-T$ space for a constant $M \approx 0.069$ in an open (solid line) and closed (dotted line) cylindrical system. The binodal curve shown with dashes. (b,c) The location of $r_i$ with respect to $M$ with constant $T/T_c=0.975$ (b) and $T$ with constant $M=0.02$ (c) for open (open symbols) and closed (filled symbols) systems, where $\phi_0 = 0.29 (\bigtriangledown) $, $0.31 (\bigcirc)$, and $0.33 (\bigtriangleup)$. In (c), the dashed lines are the binodal temperatures for each value of $\phi_0$, and the $y$-axis is on a $\log$ scale. For all figures, $R_2/R_1=5$ in the closed system.}%
\label{F_bin}%
\end{center}
\end{figure}

Delving more deeply into the requirements for a phase transition, we vary $M$ for a constant $\phi_0$ and $T$. Figure~\ref{fig_varM} shows that certain values of $M$ induce a transition, whereas others do not. In fact, there exists a critical $M_c$ that marks the lowest $M$ necessary for fluid-fluid separation. If an electric field can cause phase separation in a region of $\phi_0-T$ space \emph{above} the binodal curve, a natural question arises: what is the new stability diagram for a particular $M$? This can be constructed by holding $M$ constant and probing $\phi_0-T$ space for fluid-fluid demixing. Since the electric field breaks the symmetry of the free energy with respect to composition ($\phi_0\to 1-\phi_0$), the stability diagram is asymmetric with respect to $\phi_0-\phi_c$. Figure~\ref{fig_binodal} compares a typical stability curve for an open system, solid line, to the binodal curve, dashed line. Clearly, nonuniform fields can produce large changes to the phase diagram, in comparison with a uniform field.

Figure~\ref{fig_binodal} also includes the stability diagram for a typical closed system, dotted line, highlighting significant differences from open systems. Notably, the same $M$ induces a weaker effect and produces a smaller stability diagram when $\phi_0 < \phi_c$, and the transition can occur when $\phi_0 > \phi_c$. These differences develop as a consequence of material conservation. Since there is no infinite bath from which to draw material, accumulation of $\phi$ in high field regions leads to depletion of $\phi$ in low field regions. Moreover, the penalty in $f_{\mathrm{m}}$ grows faster than the energy gain in $f_{\mathrm{e}}$ with changing $\phi_0$ or $T$. Roughly speaking, however, the area under the stability curves for both open and closed systems increases (decreases) as $M$ increases (decreases).

Once an interface exists, several parameters control the location of $r_i$, for example $\phi_0$, $T$, $M$ and $R_2$. In general, $r_i$ increases with increasing $M$ (Fig.~\ref{fig_varM} and~\ref{fig_sigma}), decreasing $T$ (Fig.~\ref{fig_T}), increasing $R_2$ (not shown), and increasing $\phi_0$ (Fig.~\ref{fig_sigma}). The interested reader can find more specific details in Ref.~\cite{tsori2004,marcus2008,samin2009}.

\section{Phase separation dynamics}\label{tsori_dynamics}
Let's begin with a fluid-fluid mixture without an electric field and in equilibrium. When the electric field switches ``on'', the fluid mixture rearranges with time until it reaches a new equilibrium state. To describe the phase ordering dynamics, we follow the model H theoretical framework of the Hohenberg and Halperin classification~\cite{hohenberg1977}. We assume that the electrostatic  potential responds instantaneously to any changes in the composition, and supplement the model with Gauss's law to obtain
\begin{eqnarray}
\label{eq_EL_dyn1}
\frac{\partial\phi}{\partial t} + \mathbf{v}\cdot\nabla\phi &=& D\nabla^2\frac{\delta \mathcal{F}}{\delta \phi} \\
\label{eq_EL_dyn2}
\nabla\cdot(\rho \mathbf{v}) &=& 0 \\
\label{eq_EL_dyn3}
\rho\left[\frac{\partial\mathbf{v}}{\partial t}+
(\mathbf{v}\cdot\nabla)\mathbf{v}\right] &=& \eta\nabla^2\mathbf{v}-\nabla
P-\phi\nabla\frac{\partial \mathcal{F}}{\partial\phi} \\
\label{eq_EL_dyn4}
\nabla\cdot \left[\varepsilon_0\varepsilon(\phi)\nabla\psi\right] &=&0
\end{eqnarray}
where $\mathbf{v}$ is the fluid hydrodynamic velocity field, $P$ is the pressure, $\rho$ is the fluid density, and $\eta$ is fluid viscosity. The Cahn-Hilliard equation, eq.~\ref{eq_EL_dyn1}, is the continuity equation for the mixture composition. The composition changes via two mechanisms: 1) a diffusive current that depends on gradients in the chemical potential $\mu = \delta\mathcal{F}/\delta\phi$, where $D$ is the diffusivity constant, and 2) a convective current due to fluid velocity $\mathbf{v}$. Equation~\ref{eq_EL_dyn2} is the continuity equation for the fluid, while eq.~\ref{eq_EL_dyn3} is the Navier-Stokes equation that includes a chemical potential-related body force $-\phi\nabla \delta \mathcal{F}/\delta\phi$~\cite{onuki_book}. Lastly, eq.~\ref{eq_EL_dyn4} is Gauss's law, which is again coupled through $\varepsilon(\phi)$ to the other equations.

We will solve a simplified version of the model H dynamics, specifically in the over damped limit with no net fluid flow: $\mathbf{v}=0$. This is known as model B dynamics. The problem is now reduced to solving eq.~\ref{eq_EL_dyn1}, with only a diffusive current, and eq.~\ref{eq_EL_dyn4}. When $T$ decreases from above to below the critical point in the absence of an electric field, diffusion governs the exchange of material, and the fluids phase separate locally into small domains that grow in time. Most notably, this process (at late times) can be described by a characteristic length such that the domain structure at all times is self-similar when rescaled by this length~\cite{cahn1958}. Adding an electric field, as we shall see, alters this behavior.

\begin{figure}[tb]
\begin{center}
\subfigure[\label{F_dynamics_a}]{\includegraphics[keepaspectratio=true,width=0.32\textwidth]{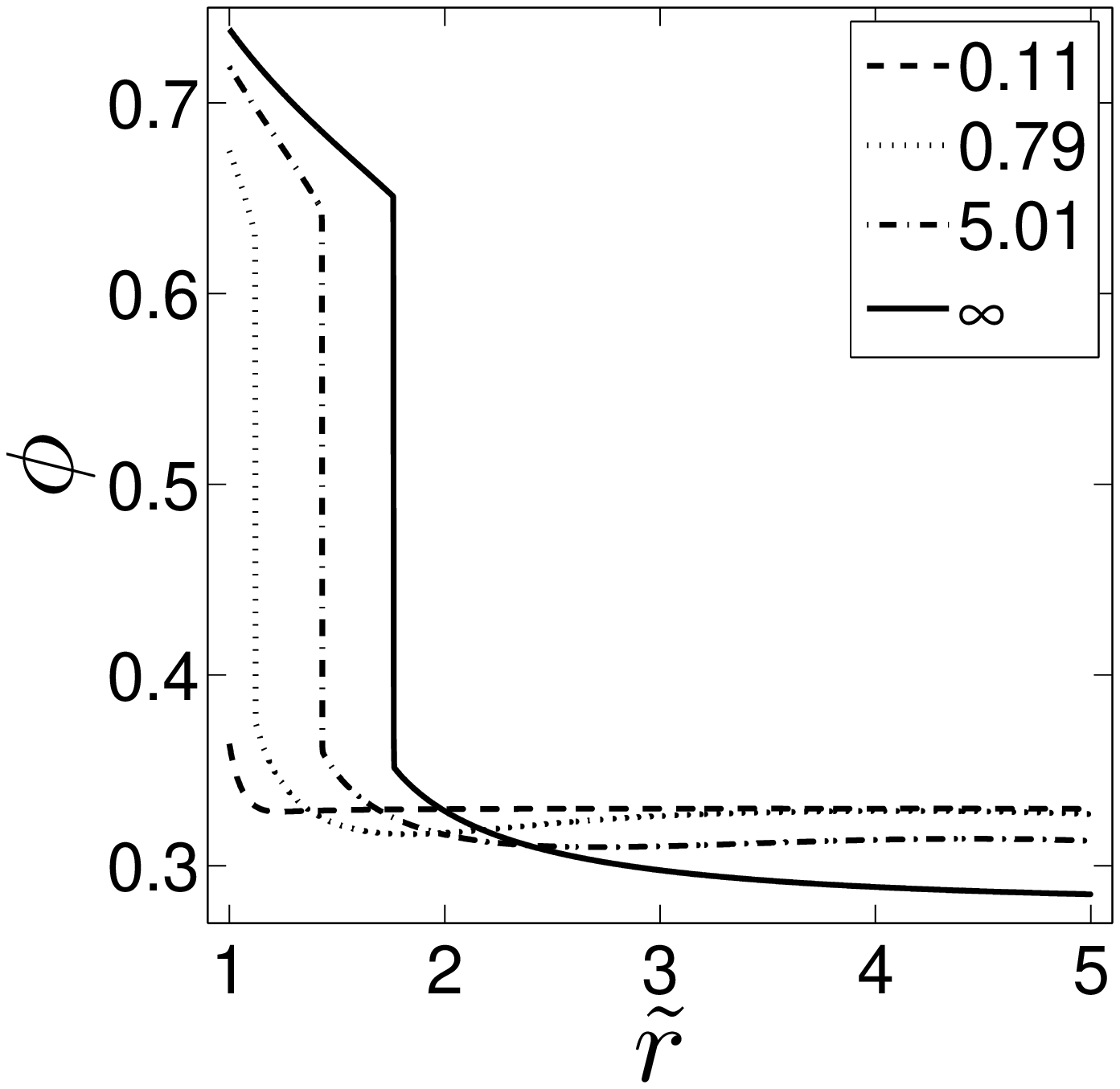}}
\subfigure[\label{F_dynamics_b}]{\includegraphics[keepaspectratio=true,width=0.32\textwidth]{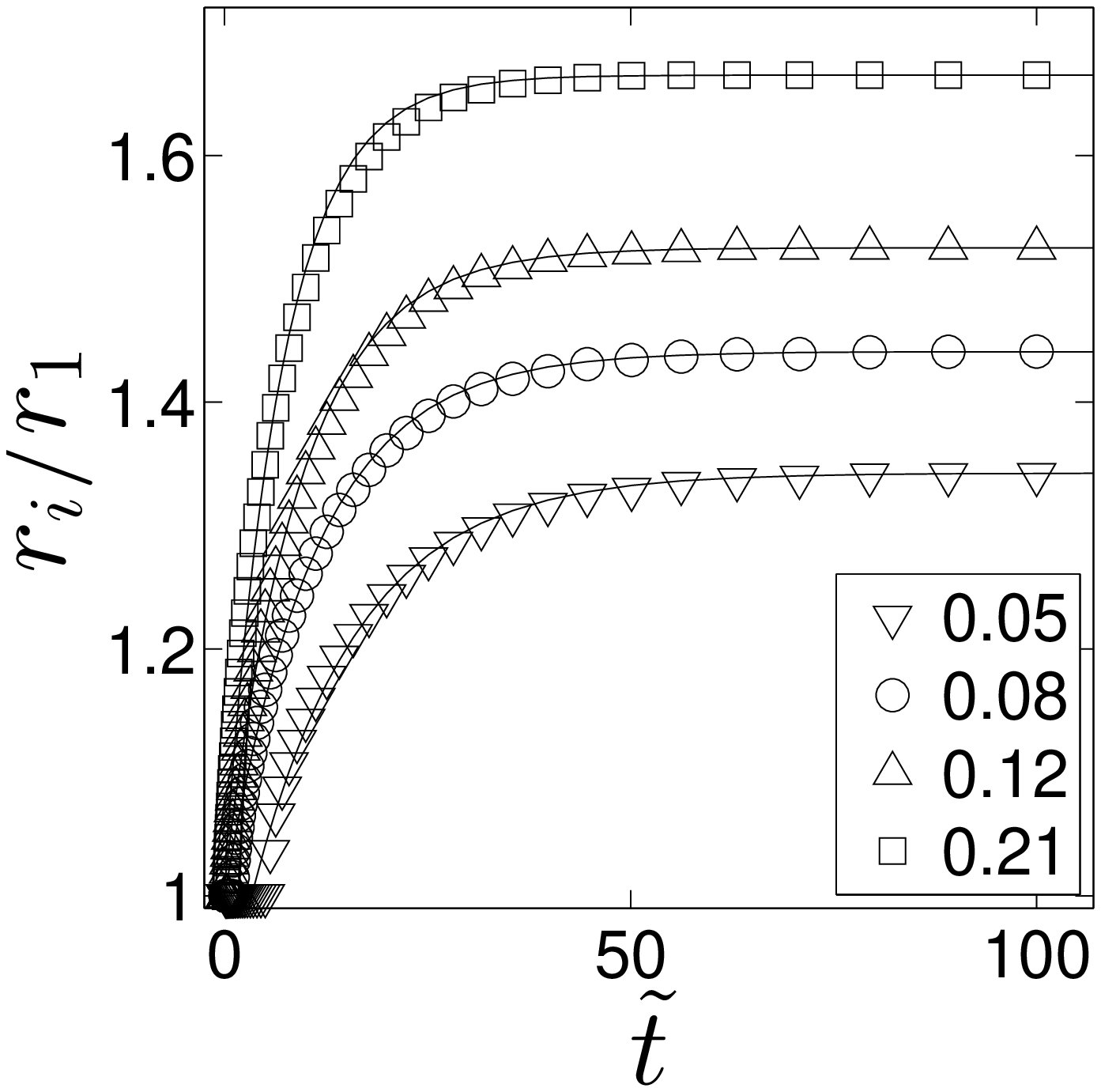}}
\subfigure[\label{F_dynamics_c}]{\includegraphics[keepaspectratio=true,width=0.32\textwidth]{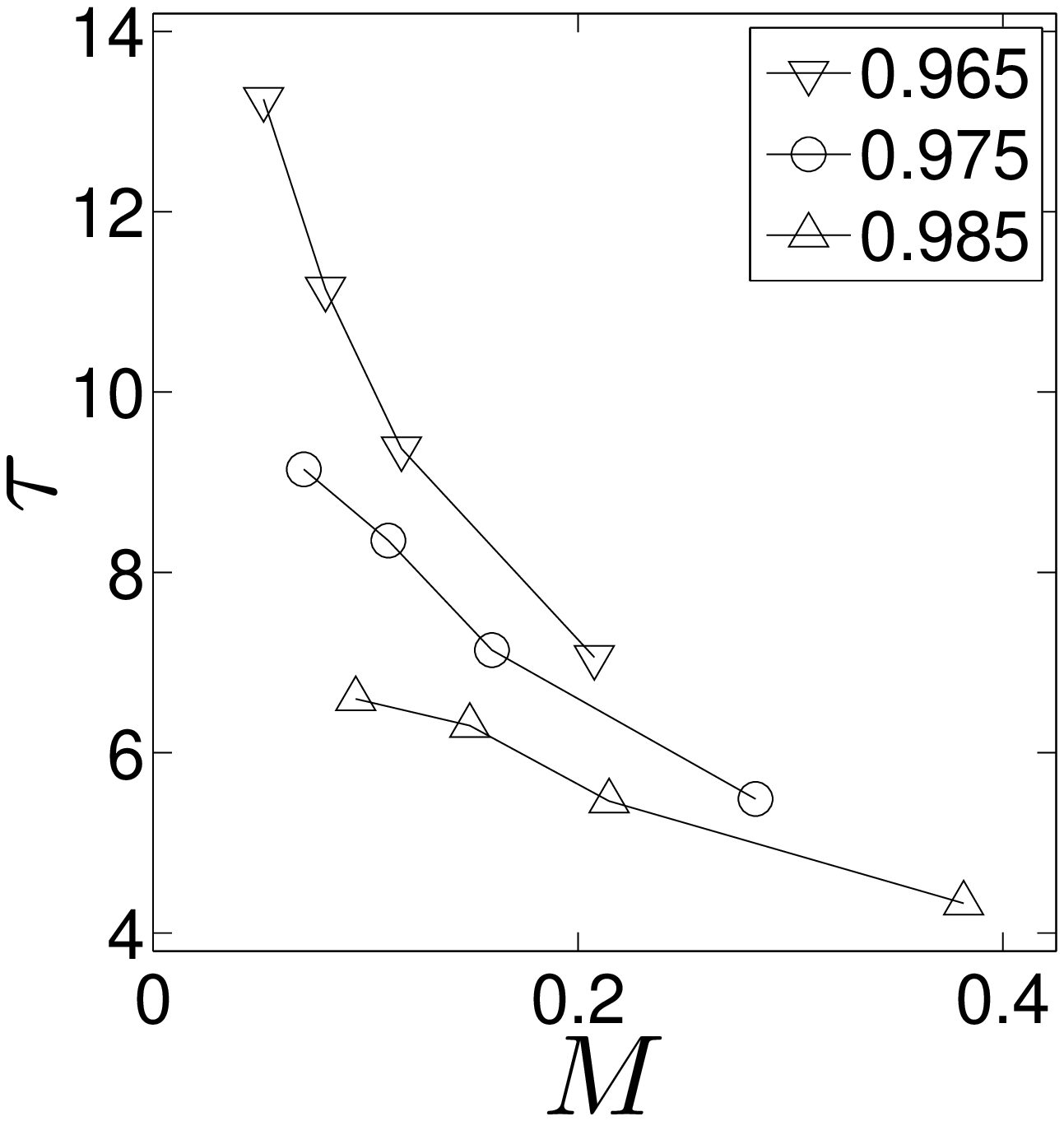}}
  \caption{\footnotesize\textsf{Movement of $r_{i}$ in time $t$. (a) $\phi(r)$ versus dimensionless distance $\tilde{r}$ at various dimensionless times $\tilde{t}$, where $\phi_0=0.33$, $T/T_c=0.965$, $M\approx 0.31$. (b) Symbols show interface location $r_i$ versus $\tilde{t}$ for various $M$, where $\phi_0=0.33$, $T/T_c=0.965$. Lines are fits to eq.~\ref{e_ri}. (c) The time constant $\tau$ as a function of $\sigma$ for various $T/T_c$. For all data, $R_2/R_1=5$.}}%
\label{F_dynamics}
\end{center}
\end{figure}

Returning to concentric cylinders in a closed system, we use the right-hand side of eq.~\ref{eq_EL_cyl} for the chemical potential in eq.~\ref{eq_EL_dyn1} and solve the radially symmetric problem with a dimensionless time $\tilde{t}=(DkT/R_1^2v_0)t$. When an electric field is turned on, material first accumulates near $r\approx R_1$. If $M>M_c$, a fluid-fluid interface emerges at $r_i(t_i) = R_1$, travels outward to larger $r$, and asymptotically reaches the long-time steady-state location $r_i(t\to\infty)=r_{i\infty}$. Figure~\ref{F_dynamics_a} shows snapshots of $\phi(r)$ in time, while the symbols in Fig.~\ref{F_dynamics_b} mark $r_i$ with time.

The behavior of $r_i$ can be approximated as an exponential relaxation
\begin{equation}
r_i(t) = r_{i\infty}+\left(r_1-r_{i\infty}\right)\exp[-(\tilde{t}-t_i)/\tau]
\label{e_ri}
\end{equation}
which contains two free parameters, the time constant for relaxation $\tau$ and the lag time for the interface to emerge $t_i$. The lines in Fig.~\ref{F_dynamics_b} show fits to the data. In general, $\tau$ decreases with increasing $M$ (Fig.~\ref{F_dynamics_c}), increasing $T$ (Fig.~\ref{F_dynamics_c}), and increasing $\phi_o$ (not shown).

In conclusion, we briefly reviewed important features of how nonuniform electric fields induce a fluid-fluid mixing-demixing phase transition. The advantage of nonuniform fields, over uniform fields, is that a phase transition can occur with only a simple dielectric difference between the fluids. Moreover, nonuniform fields can dramatically alter the transition temperature, compared with uniform fields. 

Because phase changes often lead to different material properties, the ability to easily control a phase transition often translates into the ability to easily tune material behavior. For example, turning ``on'' an electric field can cause a homogeneous fluid-fluid mixture to phase separate, which creates a refractive index mismatch and produces an optical interface. Turning the field ``off'' reverses this process. Since nonuniform fields readily occur when electrical components are small, field-induced separation may have an important technological impact.

\bibliographystyle{psp-book-har}    
\bibliography{psp-book-tsori}      

\end{document}